\documentclass[pre,preprint,showpacs,preprintnumbers,amsmath,amssymb]{revtex4}
\usepackage{psfig}

\begin{document}

\title{Accurately modeling the Internet topology}

\author{Shi Zhou}\altaffiliation[Also at ]{Department of Electronic \& Electrical Engineering,
University College London, Adastral Park Campus, Ross Building,
Martlesham Heath, Ipswich, IP5 3RE, United Kingdom.}
\author{Ra\'ul J.
Mondrag\'on}\affiliation{Department~of~Electronic~Engineering,\\
Queen~Mary~College, University~of~London,\\London, E1~4NS,
United~Kingdom. \\Email: s.zhou@adastral.ucl.ac.uk,
r.j.mondragon@elec.qmul.ac.uk}

\date{\today}

\begin{abstract}
Based on measurements of the Internet topology data, we found out
that there are two mechanisms which are necessary for the correct
modeling of the Internet topology at the Autonomous Systems (AS)
level: the Interactive Growth of new nodes and new internal links,
and a nonlinear preferential attachment, where the preference
probability is described by a positive-feedback mechanism. Based
on the above mechanisms, we introduce the Positive-Feedback
Preference (PFP) model which accurately reproduces many
topological properties of the AS-level Internet, including: degree
distribution, rich-club connectivity, the maximum degree, shortest
path length, short cycles, disassortative mixing and betweenness
centrality. The PFP model is a phenomenological model which
provides a novel insight into the evolutionary dynamics of real
complex networks.
\end{abstract}


\maketitle

\section{Introduction}

\begin{figure}[h]
\centerline{\psfig{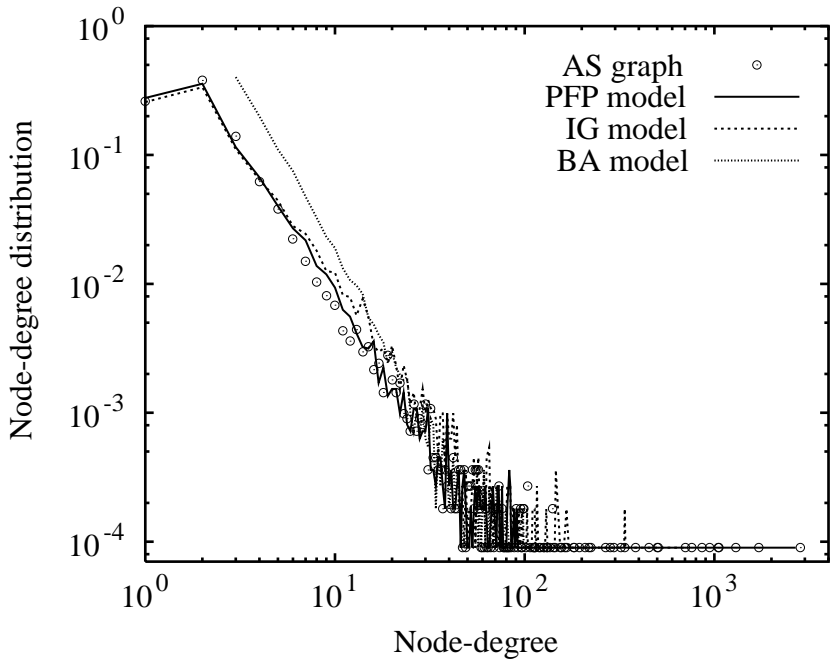}}
\caption{\label{fig:Degree}Degree distribution. The AS-level
Internet topology data used in this research is a
traceroute-derived AS graph measured in April
2002~\cite{skitterdata}.}
\end{figure}

\begin{figure}[h]
\centerline{\psfig{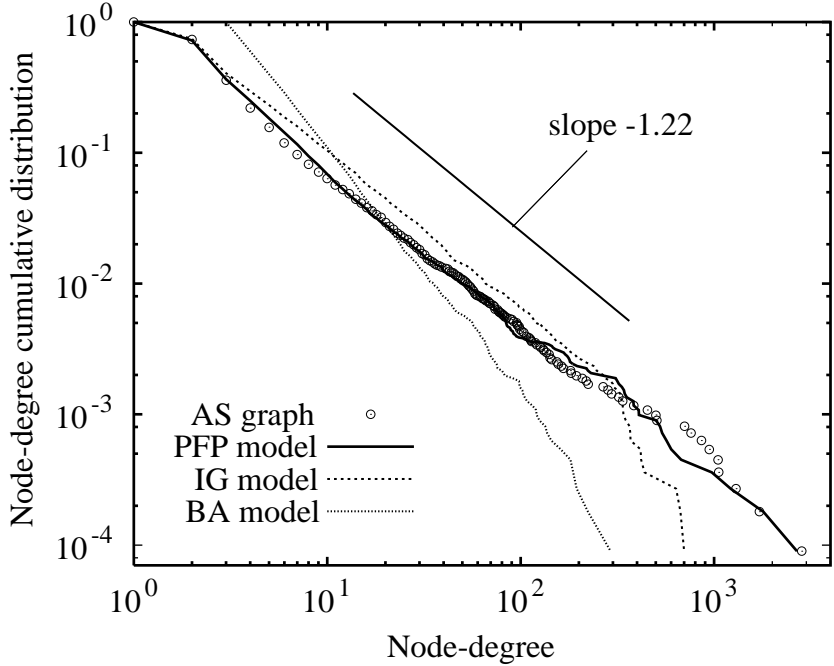}}
\caption{\label{fig:DegreeCum} The cumulative degree distribution
$P_{cmm}(k)$ of the AS graph decays as $P_{cmm}(k)\sim k^{-1.22}$,
hence the degree distribution $P(k)\sim~k^{-\gamma}$ with exponent
$\gamma\simeq 2.22$~\cite{faloutsos99}.}
\end{figure}

Recently there has been a considerable effort to understand the
topology of complex networks~\cite{strogatz01, krapivsky01,
barabasi02, albert02, bornholdt02, dorogovtsev03}. Of particular
interest are complex networks obtained from evolving mechanisms,
like  the Internet or the World Wide Web, as they are so
influential in our daily life. The degree $k$ of a node, is the
number of links which have the node as an end-point, or
equivalently, the number of nearest neighbors of the node. The
statistical distribution  of the degree $P(k)$, gives important
information of the global properties of a network and can be used
to characterize different network topologies. The Internet has
been studied in detail~\cite{faloutsos99,
 subramanian02, chen02, pastor01, vazquez02, park03, pastor04}
since the measured data~\cite{NLANR, CAIDA, oregon, michigan}
became available. Now, it is well known that the Internet can be
represented as a {\em Scale-Free} (SF) network, where the degree
distribution is a power law $P(k)\sim k^{-\gamma}$. The exponent
$\gamma$  of the Internet at the Autonomous Systems (AS) level is
approximately $2.22$ (see Fig.\,\ref{fig:Degree} and
Fig.\,\ref{fig:DegreeCum}).

Barab\'asi and Albert (BA)~\cite{barabasi99a}  showed that it is
possible to grow a network with a power-law degree distribution by
using a preferential-growth mechanism: starting with a small
random network, the system grows by attaching a new node with $m$
links to $m$ different ``old'' nodes that are already present in
the system ($m=3$ to obtain Internet-like networks); the
attachment is preferential because the probability that a new node
will connect to node $i$, with degree $k_i$, is
\begin{equation}
\Pi(i) = {k_i\over\sum_j k_j}. \label{eq:barabasi}
\end{equation}
The BA model generates networks with the power-law exponent
$\gamma=3$~\cite{barabasi99b}.

Based on the BA model, a number of evolving network models~\cite{
krapivsky01, albert02, dorogovtsev03, pastor04} have been
introduced to obtain degree distributions with other power-law
exponents. Some of these new models have been used to model the
Internet. However, a network model based solely on the
reproduction of the power-law exponent of the degree distribution
has its limitations, as it will not describe the Internet
hierarchical structure~\cite{subramanian02}. In the next section
we investigate two properties of the Internet which were not
accurately modeled by the existing models, namely the rich-club
connectivity~\cite{zhou04a} among high-degree nodes and the
maximum degree of the network. The accurate modeling of these two
properties was our motivation for developing a new network model.
In section~3 we introduce the Positive Feedback Preferential (PFP)
model, which is a phenomenological model of the AS-level Internet
topology. Section~4 presents the validation of the model and in
section~5 are the conclusions of this work.

\section{Challenges in Accurate Modeling of the Internet}

\subsection{The Rich-Club}
Scale-free networks can be grouped into assortative,
disassortative and neutral networks \cite{newman02, newman03,
 maslov04}. Social networks (e.g. the co-authorship network)
are assortative networks, in which high-degree nodes prefer to
attach to other high-degree nodes. Information networks (e.g. the
World Wide Web and the Internet) and biological networks (e.g.
protein interaction networks) have been classified as
disassortative networks, in which high-degree nodes tend to
connect with low-degree ones.

\begin{figure}[h]
\centerline{\psfig{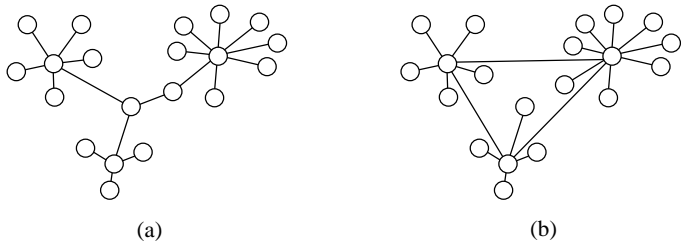}}
\caption{\label{fig:diagRichClub}Two disassortative networks. (a)
High-degree nodes are loosely interconnected. (b) High-degree
nodes are tightly interconnected.}
\end{figure}

While the AS-level Internet is disassortative~\cite{pastor01,
vazquez02}, this property does not imply that the high-degree
nodes are tightly interconnected to each other (see
Fig.~\ref{fig:diagRichClub}). One of the structural properties of
the AS-level Internet is that it contains a small number of
high-degree nodes. We call these nodes, ``rich'' nodes, and the
set containing them, the ``rich-club''. The inter-connectivity
among the club members is quantified by the rich-club
connectivity~\cite{zhou04a} which is defined as follows. The rank
$r$ of a node denotes its position on a list of all nodes sorted
in decreasing degree. If the network has  $N$ nodes then $r\in[1,
N]$. If the rich-club consists of the first $r$ nodes in the rank
list, then the rich-club connectivity $\phi(r/N)$ is defined as
the ratio of the number of links connecting the club members over
the maximum number of allowable links, $r(r-1)/2$. The rich-club
connectivity measures how well club members ``know'' each other. A
rich-club connectivity of $1$ means that all the members have a
direct link to any other member, i.e. they form a fully connected
subgraph.

\begin{figure}[h]
\centerline{\psfig{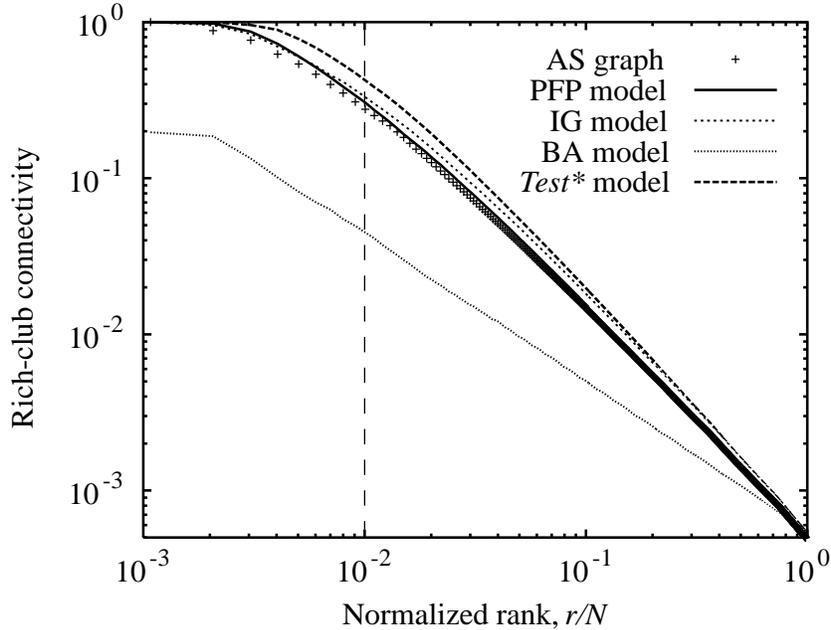}}
\caption{\label{fig:RichClub}Rich-club connectivity $\phi(r/N)$ vs
normalized rank $r/N$. The top 1\% best-connected nodes are marked
with the vertical hash line.}
\end{figure}

\begin{table}
\caption{\label{table:parameter}Network Parameters}
\begin{ruledtabular}
\begin{tabular}{cccccc}
&                           &  AS graph &  PFP model &  IG model &  BA model \\
\hline Number of nodes& $N$ & 11122     & 11122     & 11122     & 11122\\
Number of links& $L$          & 30054     & 30151     & 33349     & 33349\\
Average degree& $\langle k\rangle $  & 5.4  & 5.4&6.0& 6.0 \\
Exponent of power-law& $\gamma$ & 2.22 & 2.22&  2.22 & 3 \\
Rich-club connectivity& $\phi(r/N=0.01)$ & 0.27 & 0.30  & 0.32  & 0.045\\
Max. degree& $k_{max}$ & 2839      & 2785      & 700 & 292 \\
Degree distribution& $P(k=1)$   & 26\% & 28\%  & 26\% & 0\%\\
Degree distribution    & $P(k=2)$  & 38\% & 36\% & 34\% & 0\%\\
Degree distribution           & $P(k=3)$  & 14\% & 12\% & 11\% & 40\%\\
Characteristic path length& $l^*$ & 3.13   & 3.14      & 3.6       & 4.3 \\
Average triangle coef.& $\langle k_t\rangle $   & 12.7      & 12        & 10.4      & 0.1 \\
Max. triangle coef.& $k_{t-max}$     & 7482      & 8611      & 4123      & 64 \\
Average quadrangle coef.& $\langle k_q\rangle $   & 277       & 247       & 105.4     &  1.3\\
Max. quadrangle coef.& $k_{q-max}$      & 9648      & 9431      & 8780      & 527 \\
Average $k_{nn}$ & $\langle k_{nn}\rangle $ & 660   &482  &  103 & 20 \\
Average betweenness& $\langle {\cal C}_B^*\rangle $ & 4.13   & 4.14 & 4.6 & 5.3\\
Max. betweenness& ${{\cal C}_B^*}_{-max}$ & 3237   & 3419 & 1002 & 1064\\
\end{tabular}
\end{ruledtabular}
\end{table}

Fig.\,\ref{fig:RichClub} shows the rich-club connectivity as a
function of the  rank normalized by the number of nodes. It is
clear that in the AS graph the high-degree nodes are tightly
interconnected. The top 1\% best-connected nodes of the AS graph
have 27\% of the possible interconnections, compared with only
4.5\% obtained from a network topology generated using the BA
model which has the same number of nodes and slightly larger
number of links as the AS graph (see table I).

The rich-club consists of highly connected nodes, which are well
interconnected between each other and the average hop distance
among the club members is very small (1 to 2 hops). The rich-club
is a ``super'' traffic hub of the network and the disassortative
mixing property ensures that peripheral nodes are always near the
hub. These two structural properties together contribute to the
routing efficiency of the network. An Internet model that does not
reproduces the properties of the rich-club will underestimate the
actual network's routing efficiency (shortest path length) and
routing flexibility (alternative reachable paths), and also, it
will  overestimate the network robustness under
node-attack~\cite{zhou04b}.

\subsubsection*{The Interactive Growth Model}
The BA model is based solely on the attachment of new nodes.
However the appearance of new internal links among old nodes has
also been observed in the evolution of the
Internet~\cite{pastor01, vazquez02}. During the last few years,
researchers have proposed a number of Internet models using the
appearance of new internal links, such as Dorogovtsev and Mendes'
model~\cite{dorogovtsev00}, Bu and Towsley's Generalized Linear
Preference (GLP) model~\cite{bu02}, Bianconi~{\sl et~al} 's
Generalized Network Growth (GNG) model~\cite{bianconi03},
Caldarelli~{\sl et~al} 's model~\cite{caldarelli03} and the
Interactive Growth (IG) model~\cite{zhou03c}. In addition to the
appearance of new internal links, these models have also used
different preference schemes to capture selected properties of the
Internet.

Here we revisit the Interactive Growth (IG) model as it is the
precursor of the Positive Feedback Preference model and the IG
model provides a possible way to reproduce both the power-law
degree distribution and the rich-club connectivity of the AS
graph.  The IG model generates a network using the {\em
Interactive Growth}, where new internal links start from the host
nodes, which are the old nodes  that new nodes are attached to.
The IG model starts with a small random network, at each time
step,
\begin{itemize}
\item  with probability $p\in(0, 1)$, a new node is attached to
one host node and two new internal links appear between the host
node and two other old nodes (peer nodes); \item with probability
$1-p$, a new node is attached to two host nodes and one new
internal link appears between one of the host nodes and a peer
node.
\end{itemize}
In the actual Internet, new nodes bring new traffic load to its
host nodes. This results in both the increase of traffic volume
and the change of traffic pattern around host nodes and triggers
the addition of new links connecting host nodes to peer nodes in
order to balance network traffic and optimize network performance.
From numerical simulations, we found that when $p=0.4$ the
Interactive Growth also satisfies the following two
characteristics observed~\cite{chen02,pastor01, vazquez02,park03}
in the Internet measurements. Firstly, the majority of new nodes
are added to the system by attaching them to one or two old nodes
($m\le2$). Secondly the degree distribution of the AS graph is not
a strict power-law as it has more nodes with degree two than nodes
with degree one ($P(2)=38\%>P(1)=26\%$, see
Table~\ref{table:parameter}). The IG model uses the BA model's
linear preference of Eq.\,(\ref{eq:barabasi}) in the attachment of
new nodes and the appearance of new internal links. As shown in
Fig.\,\ref{fig:Degree}, Fig.\,\ref{fig:DegreeCum},
Fig.\,\ref{fig:RichClub} and Table~\ref{table:parameter}, the IG
model closely resembles both the power-law degree distribution and
the rich-club connectivity of the AS graph.

\subsection{Maximum Degree}

\begin{figure}[tbh]
\centerline{\psfig{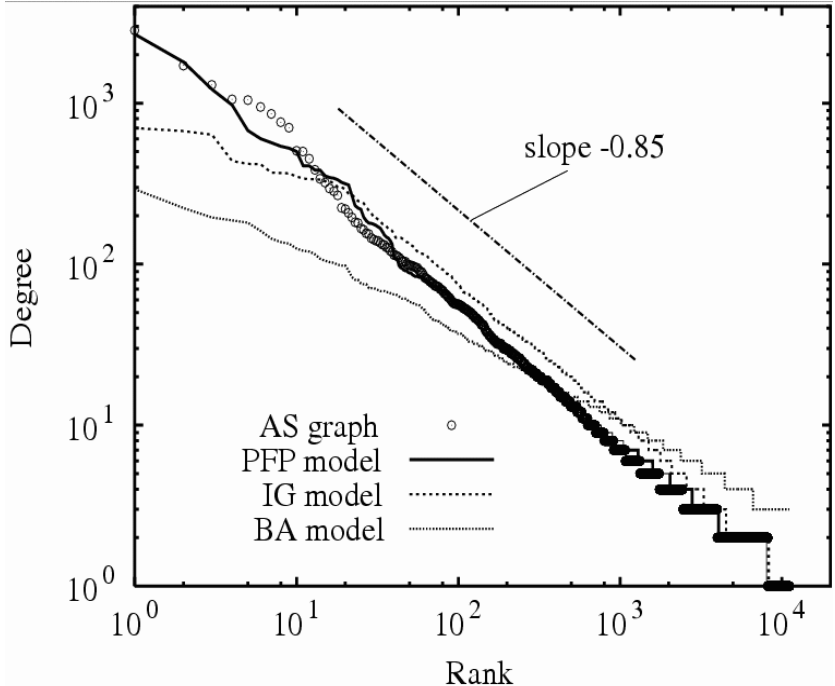}}
\caption{\label{fig:DegreeRank}Node degree $k$ vs rank $r$.}
\end{figure}

The IG model still has its limitations. The maximum node degree
$k_{max}$ present in the AS graph is nearly a quarter of the
number of nodes ($k_{max}\simeq N/4$) and is significantly larger
than the maximum degree obtained by the IG and BA models using
linear preferential attachment (see Table~\ref{table:parameter}).
To overcome this shortfall, it is possible to favor high-degree
nodes by using the nonlinear preferential
probability~\cite{dorogovtsev00,krapivsky00}
\begin{equation}
\Pi(i)={k_i^\alpha\over\sum_{j}
k_j^\alpha},~~\alpha>1.\label{eq:NonLineal}
\end{equation}

To examine the above nonlinear preference, here we study a
so-called \emph{Test*} model, which is a modification of the IG
model. The \emph{Test*} model uses the same Interactive Growth
mechanism as the IG model, but it does not use the linear
preference given by Eq.\,(\ref{eq:barabasi}), instead it uses the
nonlinear preference given by Eq.\,(\ref{eq:NonLineal}). Numerical
experiments showed that, when $\alpha=1.15\pm0.01$, the
\emph{Test*} model generates networks with the maximum degree
similar to the AS graph. However, as shown in
Fig.~\ref{fig:RichClub}, the rich-club connectivity produced by
the \emph{Test*} model deviates from the AS graph. For example,
the 1\% best connected nodes of the \emph{Test*} model have 42\%
allowable interconnections   compared with 27\% of the AS graph.

\section{Positive-Feedback Preference Model}

Based on the Internet-history data, Pastor-Satorras~{\sl
et~al}~\cite{pastor01}  and V\'azquez~{\sl et~al}~\cite{vazquez02}
measured that the probability that  a new node links with a
low-degree old node follows the linear preferential attachment
given by Eq.~(\ref{eq:barabasi}). Whereas Chen~{\sl
et~al}~\cite{chen02} reported that high-degree nodes have a
stronger ability of acquiring new links than predicted by
Eq.~(\ref{eq:barabasi}). The Internet-history data also show that
at early times, the degree of node increases very slowly; later
on, the degree grows more and more rapidly. Taking into account
these observations, we modified the IG model by using the
nonlinear preferential attachment
\begin{equation} \Pi(i) =
{k_i^{1+\delta\log_{10}{k_i}}\over\sum_j
k_j^{1+\delta\log_{10}{k_j}}},~~\delta\in[0, 1]. \label{eq:PFP}
\end{equation}
We call this the
Positive-Feedback Preference (PFP) model. From numerical
simulations, we found that $\delta=0.048$ produces the best
result. (It is interesting to notice that for $\delta=0.048$ and
the maximum degree $k_{max}=2839$ as measured on the AS graph, the
exponent function of $1+\delta\log_{10}{k_{max}}\simeq1.166$,
which is close to the value of $\alpha$ used in the \emph{Test*}
model).

We also refine the Interactive Growth mechanism. The PFP model
starts with a small random network, at each time step,
\begin{itemize}
\item  with probability $p\in[0, 1]$, a new node is attached to
one host node;  and at the same time one new internal link appears
between the host node and a peer node;

\item with probability $q\in[0, 1-p]$, a new node is attached to
one host node; and at the same time two new internal links appear
between the host node and two peer nodes;

\item with probability $1-p-q$, a new node is attached to two host
nodes; and at the same time one new internal link appears between
one of the host nodes and one peer node;
\end{itemize}
When $p=0.3$ and $q=0.1$, the generated PFP network has the same
ratio of nodes to links as in the AS graph (see
Table~\ref{table:parameter}). Eq.\,(\ref{eq:PFP}) is used in
choosing host nodes and peer nodes.

\begin{figure}[h]
\begin{minipage}[t]{80mm}
\centerline{\psfig{figure=figure6_kfunctions.ai,width=8cm}}
\caption{\label{fig:DegreeFunctions}Three degree functions: $k$,
$k^{\alpha}$ with $\alpha=1.15$ and $k^{1+\delta\log_{10}{k}}$
with $\delta=0.048$.}
\end{minipage}
\hspace{\fill}
\begin{minipage}[t]{80mm}
\centerline{\psfig{figure=figure7_k_age.ai,width=8cm}}
\caption{\label{fig:DegreeTime}Degree growth of a node. }
\end{minipage}
\end{figure}

The PFP model satisfies Pastor-Sartorras~{\sl et~al},
V\'azquez~{\sl et~al} and Chen~{\sl et~al} 's observations. For
low-degree nodes, the  preferential attachment is approximated by
Eq.~(\ref{eq:barabasi}). For high-degree nodes, the preferential
attachment increases as a nonlinear function of the node degree
(see Fig.~\ref{fig:DegreeFunctions}).  Hence,
 as the time passes by, the rate of degree growth in the
PFP model is faster than in the IG model and the BA model (see
Fig.~\ref{fig:DegreeTime}).

\section{Model Validation}
The validation was done by comparing the AS
graph~\cite{skitterdata} with networks generated by the PFP model,
the IG model and the BA model. For each model, ten different
networks were generated and averaged. The networks had the same
number of nodes and similar numbers of links as the AS graph (see
Table~\ref{table:parameter}).

\subsection{Degree Distribution, Rich-Club Connectivity and Maximum Degree}
The PFP model produces networks that closely matches the degree
distribution (see Fig.\,\ref{fig:Degree} and
Fig.\,\ref{fig:DegreeCum}), the rich-club connectivity (see
Fig.~\ref{fig:RichClub}) and the maximum degree (see Table~I) of
the AS graph. Also the networks generated using the PFP model have
the same power-law relationship between degree and rank, $k\sim
r^{-0.85}$ as the AS graph (see Fig.~\ref{fig:DegreeRank}).
In certain respect the accuracy of the PFP model to reproduce these properties   is not a surprise.
After all, the model was designed to match these properties.

\subsection{Shortest-Path Length}

\begin{figure}[htb]
\centerline{\psfig{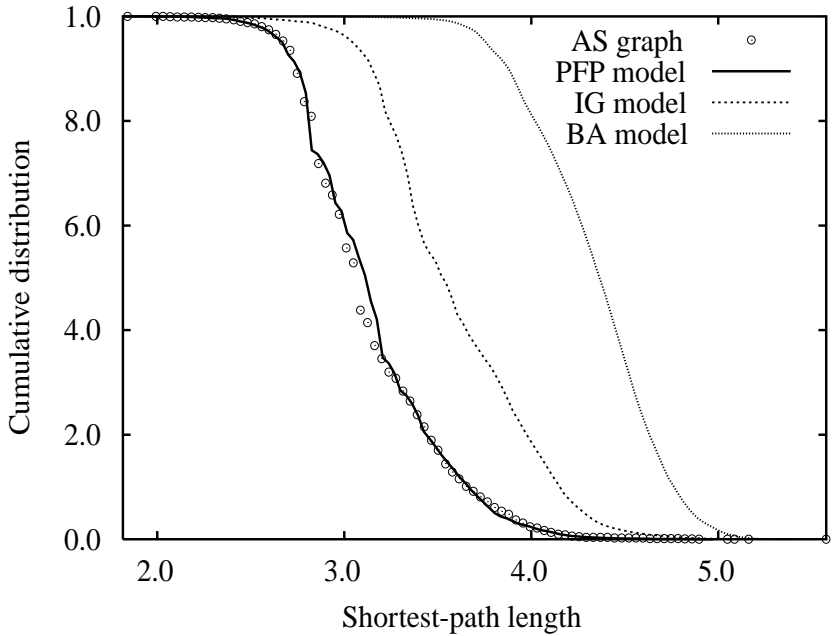}}
\caption{\label{fig:LengthDistribution}Cumulative distribution of
average shortest-path length.}
\end{figure}

\begin{figure}[htb]
\centerline{\psfig{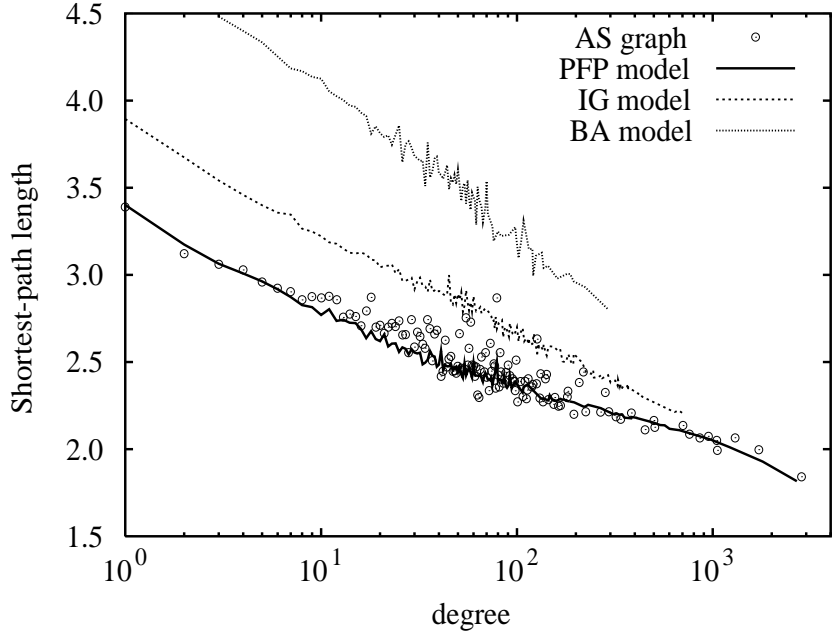}}
\caption{\label{fig:LengthDegree}Correlation between average
shortest-path length $l$ and degree, where $l$ is the average over
nodes with the same degree.}
\end{figure}

The average shortest-path length $l$, of a node is defined as the
average of the shortest-paths  from the node to all other nodes in
the network. Fig.\,\ref{fig:LengthDistribution} and
Fig.\,\ref{fig:LengthDegree} show that the PFP model reproduces
the cumulative distribution of average shortest-path length and
the correlation between average shortest-path length and degree of
the AS graph.

The characteristic path length $l^*$, of a network is the average
of the shortest-paths over all pairs of nodes. The characteristic
path length indicates the network overall routing efficiency. The
AS graph is a small-world network~\cite{watts99} because the
characteristic path length is very small compared with the network
size. Table~\ref{table:parameter} shows that the AS graph and the
networks obtained from the PFP model have nearly the same
characteristic path length.

\subsection{Short Cycles}
Cycles~\cite{bianconi03, bianconi03a} encode the redundant
information in the network structure. The number of short cycles
(triangles and quadrangles) is a  relevant property because the
multiplicity of paths between any two nodes increases with the
density of short cycles (note that an alternative path between
two nodes can be longer than their shortest-path).
The triangle
coefficient $k_t$, is defined as the number of triangles that a
node shares. Similarly the quadrangle coefficient $k_q$, is the
number of quadrangles that a node has.

\begin{figure}[h]
\centerline{\psfig{figure=figure10_ptri.ai,width=110mm}}
\caption{\label{fig:PTriCum}Cumulative distribution of triangle
coefficient.}\vspace{15mm}
\centerline{\psfig{figure=figure11_prect.ai,width=110mm}}
\caption{\label{fig:PRectCum}Cumulative distribution of quadrangle
coefficient.}
\end{figure}

\begin{figure}[h]
\centerline{\psfig{figure=figure12_tri_k.ai,width=110mm}}
\caption{\label{fig:TriDegree} Correlation between triangle
coefficient $k_t$ and degree, where $k_t$ is the average over
nodes with the same degree.}\vspace{15mm}
\centerline{\psfig{figure=figure13_rect_k.ai,width=110mm}}
\caption{\label{fig:RectDegree}Correlation between quadrangle
coefficient $k_q$ and degree, where $k_q$ is the average over
nodes with the same degree.}
\end{figure}

Table~\ref{table:parameter} shows the AS graph and the networks
generated using the PFP model have higher densities of short
cycles ($\langle k_t\rangle $ and $\langle k_q\rangle $) than
networks generated using the IG model and the BA model.
Fig.\,\ref{fig:PTriCum} and Fig.\,\ref{fig:PRectCum} show that the
AS graph and the networks obtained from the PFP model have similar
cumulative distributions of short cycles.
Fig.\,\ref{fig:TriDegree} and Fig.\,\ref{fig:RectDegree} show that
the PFP networks exhibit similar correlations between short cycles
and degree as in the AS graph.

Notice that the clustering coefficient $c$ of a node can be
expressed as a function of the node's degree $k$ and triangle
coefficient $k_t$,
\begin{equation}
c={k_t\over k(k-1)/2}.
\end{equation}
The reason we study short cycles instead of clustering
coefficient is that short cycles have the advantage of providing
neighbor clustering information of nodes with different degrees.

\subsection{Disassortative Mixing}

The Internet exhibits the disassortative mixing
behavior~\cite{newman03, pastor01,vazquez02,maslov04}, where on
average, high-degree nodes tend to connect to peripheral nodes
with low degrees. A network's mixing pattern is identified by the
conditional probability $p_c(k'|k)$ that a link connects a node
with degree $k$ to a node with degree $k'$. This conditional
probability can be indicated~\cite{pastor01,vazquez02} by
$k_{nn}$, the nearest-neighbors average degree  of a node with
degree $k$.

\begin{figure}[h]
\centerline{\psfig{figure=figure14_pknncum.ai,width=11cm}}
\caption{\label{fig:PKnnCum}Cumulative distribution of
nearest-neighbors average degree.}\vspace{15mm}
\centerline{\psfig{figure=figure15_knn_k.ai,width=11cm}}
\caption{\label{fig:KnnDegree}Correlations between
nearest-neighbors average degree $k_{nn}$ and degree, where
$k_{nn}$ is the average over nodes with the same degree.}
\end{figure}

Fig.\,\ref{fig:PKnnCum} and Table~\ref{table:parameter} show that
on average the nearest-neighbors average degree of a node in the
AS graph and the PFP networks is significantly larger than that in
the IG and BA networks. Fig.\,\ref{fig:KnnDegree} shows that the
PFP model closely reproduces the negative correlation between
nearest-neighbors average degree and node degree of the AS graph
and therefore exhibits similar disassortative mixing as the AS
graph.

\subsection{Betweenness Centrality}

On a network, there are nodes that are more prominent because they
are highly used when transferring information. A way to measure
this ``importance'' is by using the concept of node \emph{
betweenness centrality} which is defined as follows. Given a
source node $s$ and a destination node $d$, the number of
different shortest-paths from $s$ to $d$ is $g(s,d)$. The number
of shortest-paths that contain the node $w$ is $g(w;s,d)$. The
proportion of shortest-paths, from $s$ to $d$, which contain node
$w$ is $p_{s,d}(w) = {g(w;s,d)/g(s,d)}$. The betweenness
centrality of node $w$ is defined~\cite{goh01,Holme02a} as
\begin{equation}
{\cal C}_B(w) = \sum_{s}\sum_{d\ne s}p_{s,d}(w),
\label{eq:betwCentNode}
\end{equation}
where the sum is over all possible pairs of nodes with $s\ne d$.
The betweenness centrality measures the proportion of shortest
paths which visit a certain node. If all pairs of nodes of a
network communicate at the same rate,  the betweenness centrality
estimates the node's capacity needed for a  free-flow
status~\cite{goh01}. A node with a large ${\cal C}_B$ is
``important'' because it carries a large traffic load. If this
node fails or gets congested, the consequences to the network
traffic can be drastic \cite{Holme02a}. Here the betweenness
centrality is normalized by the number of nodes and denoted as
${\cal C}_B^*$. The average of the (normalized) betweenness
centrality in a network ${\langle {\cal C}_B^*\rangle
}=l^*+1$~\cite{Holme02a}, where $l^*$ is the network's
characteristic path length.

\begin{figure}[htb]
\centerline{\psfig{figure=figure16_pbcum.ai,width=110mm}}
\caption{\label{fig:BDistribution}Cumulative distribution of
betweenness centrality, $P_{cum}({\cal C}_B^*)$. } \vspace{15mm}
\centerline{\psfig{figure=figure17_b_k.ai,width=110mm}}
\caption{\label{fig:BDegree}Correlations between betweenness
centrality ${\cal C}_B^*$ and degree, where ${\cal C}_B^*$ is the
average over nodes with the same degree.}
\end{figure}

Fig.\,\ref{fig:BDistribution} shows that the cumulative
distribution of betweenness centrality $P_{cum}({\cal C}_B^*)$ of
the networks exhibit similar power-law behaviors characterized by
slope $-1.1$, hence $P({\cal C}_B^*)\sim({\cal
C}_B^*)^{-2.1}$~\cite{pastor01,vazquez02}. However as shown in
Table~\ref{table:parameter}, the maximum value of the betweenness
centrality, ${{\cal C}_B^*}_{-max}$, for the AS graph and the PFP
model are significantly larger than that for the IG model and the
BA model. Fig.\,\ref{fig:BDegree} shows that only the PFP model
closely matches the correlation between betweenness centrality and
degree of the AS graph.

\section{ Conclusions and Discussion}

In summary, the PFP model accurately reproduces many of the
topological properties measured in the Internet at the AS level.
The model is based on two growth mechanisms which are the
nonlinear positive-feedback preferential attachment combined with
the Interactive Growth of new nodes and new internal links. Both
the mechanisms are based on (and supported by) the observations on
the Internet history data.

The positive-feedback preference means that, as a node acquires
new links, the node's relative advantage when competing for more
links increases as a non-linear feed-back loop. This implies  the
inequality in the link-acquiring ability between rich nodes and
non-rich nodes increases as the network evolves. Rich nodes, not
only become richer, they become disproportionately richer. While
our initial motivation was to create a model that can accurately
reproduce the rich-club connectivity and the maximum degree of the
AS graph, the PFP model actually captures other properties as
well. Further studies are needed to explain why the Internet
growth seems to follow the non-linear preferential attachment
given by the PFP model and what are the consequences of this
growth mechanism for the future of the Internet. This research
provides an insight into the basic mechanisms that could be
responsible for the evolving topology of complex networks.

Finally, the validation of the model was not conducted with
measurement data based on the BGP-tables, but with the
traceroute-derived AS graph, which is regarded as a more realistic
and reliable measurement of the Internet~\cite{hyun03}.

\section{Acknowledgments}
The authors would like to thank the referees for their useful
comments. This work is funded by the U. K. Engineering and
Physical Sciences Research Council (EPSRC) under grant no.
GR-R30136-01.


\begin{thebibliography}{36}
\expandafter\ifx\csname
natexlab\endcsname\relax\def\natexlab#1{#1}\fi
\expandafter\ifx\csname bibnamefont\endcsname\relax
  \def\bibnamefont#1{#1}\fi
\expandafter\ifx\csname bibfnamefont\endcsname\relax
  \def\bibfnamefont#1{#1}\fi
\expandafter\ifx\csname citenamefont\endcsname\relax
  \def\citenamefont#1{#1}\fi
\expandafter\ifx\csname url\endcsname\relax
  \def\url#1{\texttt{#1}}\fi
\expandafter\ifx\csname
urlprefix\endcsname\relax\def\urlprefix{URL }\fi
\providecommand{\bibinfo}[2]{#2}
\providecommand{\eprint}[2][]{\url{#2}}

\bibitem[{\citenamefont{Strogatz}(2001)}]{strogatz01}
\bibinfo{author}{\bibfnamefont{S.~H.} \bibnamefont{Strogatz}},
  \bibinfo{journal}{Nature (London)} \textbf{\bibinfo{volume}{410}},
  \bibinfo{pages}{268} (\bibinfo{year}{2001}).

\bibitem[{\citenamefont{Krapivsky and Redner}(2001)}]{krapivsky01}
\bibinfo{author}{\bibfnamefont{P.~L.} \bibnamefont{Krapivsky}}
  \bibnamefont{and} \bibinfo{author}{\bibfnamefont{S.}~\bibnamefont{Redner}},
  \bibinfo{journal}{Phys. Rev. E} \textbf{\bibinfo{volume}{63}},
  \bibinfo{pages}{066123} (\bibinfo{year}{2001}).

\bibitem[{\citenamefont{Barab\'asi}(2002)}]{barabasi02}
\bibinfo{author}{\bibfnamefont{A.~L.} \bibnamefont{Barab\'asi}},
  \emph{\bibinfo{title}{Linked: The New Science of Networks}}
  (\bibinfo{publisher}{Perseus Publishing}, \bibinfo{year}{2002}).

\bibitem[{\citenamefont{Albert and Barab\'asi}(2002)}]{albert02}
\bibinfo{author}{\bibfnamefont{R.}~\bibnamefont{Albert}} \bibnamefont{and}
  \bibinfo{author}{\bibfnamefont{A.~L.} \bibnamefont{Barab\'asi}},
  \bibinfo{journal}{Rev. Mod. Phys.} \textbf{\bibinfo{volume}{74}},
  \bibinfo{pages}{47} (\bibinfo{year}{2002}).

\bibitem[{\citenamefont{Bornholdt and Schuster}(2002)}]{bornholdt02}
\bibinfo{author}{\bibfnamefont{S.}~\bibnamefont{Bornholdt}} \bibnamefont{and}
  \bibinfo{author}{\bibfnamefont{H.~G.} \bibnamefont{Schuster}},
  \emph{\bibinfo{title}{Handbook of Graphs and Networks - From the Genome to
  the Internet}} (\bibinfo{publisher}{Wiley-VCH}, \bibinfo{address}{Weinheim
  Germany}, \bibinfo{year}{2002}).

\bibitem[{\citenamefont{Dorogovtsev and Mendes}(2003)}]{dorogovtsev03}
\bibinfo{author}{\bibfnamefont{S.~N.} \bibnamefont{Dorogovtsev}}
  \bibnamefont{and} \bibinfo{author}{\bibfnamefont{J.~F.~F.}
  \bibnamefont{Mendes}}, \emph{\bibinfo{title}{Evolution of Networks - From
  Biological Nets to the Internet and WWW}} (\bibinfo{publisher}{Oxford
  University Press}, \bibinfo{year}{2003}).

\bibitem[{\citenamefont{Faloutsos et~al.}(1999)\citenamefont{Faloutsos,
  Faloutsos, and Faloutsos}}]{faloutsos99}
\bibinfo{author}{\bibfnamefont{M.}~\bibnamefont{Faloutsos}},
  \bibinfo{author}{\bibfnamefont{P.}~\bibnamefont{Faloutsos}},
  \bibnamefont{and}
  \bibinfo{author}{\bibfnamefont{C.}~\bibnamefont{Faloutsos}},
  \bibinfo{journal}{Comput. Commun. Rev.} \textbf{\bibinfo{volume}{29}},
  \bibinfo{pages}{251} (\bibinfo{year}{1999}).

\bibitem[{\citenamefont{Subramanian et~al.}(2002)\citenamefont{Subramanian,
  Agarwal, Rexford, and Katz}}]{subramanian02}
\bibinfo{author}{\bibfnamefont{L.}~\bibnamefont{Subramanian}},
  \bibinfo{author}{\bibfnamefont{S.}~\bibnamefont{Agarwal}},
  \bibinfo{author}{\bibfnamefont{J.}~\bibnamefont{Rexford}}, \bibnamefont{and}
  \bibinfo{author}{\bibfnamefont{R.~H.} \bibnamefont{Katz}}, in
  \emph{\bibinfo{booktitle}{Proc. of IEEE {INFOCOM} 2002}}
  (\bibinfo{year}{2002}), pp. \bibinfo{pages}{618--627}.

\bibitem[{\citenamefont{Chen et~al.}(2002)\citenamefont{Chen, Chang, Govindan,
  Jamin, Shenker, and Willinger}}]{chen02}
\bibinfo{author}{\bibfnamefont{Q.}~\bibnamefont{Chen}},
  \bibinfo{author}{\bibfnamefont{H.}~\bibnamefont{Chang}},
  \bibinfo{author}{\bibfnamefont{R.}~\bibnamefont{Govindan}},
  \bibinfo{author}{\bibfnamefont{S.}~\bibnamefont{Jamin}},
  \bibinfo{author}{\bibfnamefont{S.~J.} \bibnamefont{Shenker}},
  \bibnamefont{and}
  \bibinfo{author}{\bibfnamefont{W.}~\bibnamefont{Willinger}}, in
  \emph{\bibinfo{booktitle}{Proc. of IEEE {INFOCOM} 2002}}
  (\bibinfo{year}{2002}), pp. \bibinfo{pages}{608--617}.

\bibitem[{\citenamefont{Pastor-Satorras
  et~al.}(2001)\citenamefont{Pastor-Satorras, V\'azquez, and
  Vespignani}}]{pastor01}
\bibinfo{author}{\bibfnamefont{R.}~\bibnamefont{Pastor-Satorras}},
  \bibinfo{author}{\bibfnamefont{A.}~\bibnamefont{V\'azquez}},
  \bibnamefont{and}
  \bibinfo{author}{\bibfnamefont{A.}~\bibnamefont{Vespignani}},
  \bibinfo{journal}{Phys. Rev. Lett.} \textbf{\bibinfo{volume}{87}},
  \bibinfo{pages}{258701} (\bibinfo{year}{2001}).

\bibitem[{\citenamefont{V\'azquez et~al.}(2002)\citenamefont{V\'azquez,
  Pastor-Satorras, and Vespignani}}]{vazquez02}
\bibinfo{author}{\bibfnamefont{A.}~\bibnamefont{V\'azquez}},
  \bibinfo{author}{\bibfnamefont{R.}~\bibnamefont{Pastor-Satorras}},
  \bibnamefont{and}
  \bibinfo{author}{\bibfnamefont{A.}~\bibnamefont{Vespignani}},
  \bibinfo{journal}{Phys. Rev. E} \textbf{\bibinfo{volume}{65}},
  \bibinfo{pages}{066130} (\bibinfo{year}{2002}).

\bibitem[{\citenamefont{Park et~al.}(2003)\citenamefont{Park, Khrabrov,
  Pennock, Lawrence, Giles, and Ungar}}]{park03}
\bibinfo{author}{\bibfnamefont{S.~T.} \bibnamefont{Park}},
  \bibinfo{author}{\bibfnamefont{A.}~\bibnamefont{Khrabrov}},
  \bibinfo{author}{\bibfnamefont{D.~M.} \bibnamefont{Pennock}},
  \bibinfo{author}{\bibfnamefont{S.}~\bibnamefont{Lawrence}},
  \bibinfo{author}{\bibfnamefont{C.~L.} \bibnamefont{Giles}}, \bibnamefont{and}
  \bibinfo{author}{\bibfnamefont{L.~H.} \bibnamefont{Ungar}}, in
  \emph{\bibinfo{booktitle}{Proc. of IEEE INFOCOM 2003}}
  (\bibinfo{year}{2003}), vol.~\bibinfo{volume}{3}, pp.
  \bibinfo{pages}{2144--2154}.

\bibitem[{\citenamefont{Pastor-Satorras and Vespignani}(2004)}]{pastor04}
\bibinfo{author}{\bibfnamefont{R.}~\bibnamefont{Pastor-Satorras}}
  \bibnamefont{and}
  \bibinfo{author}{\bibfnamefont{A.}~\bibnamefont{Vespignani}},
  \emph{\bibinfo{title}{Evolution and Structure of the Internet - A Statistical
  Physics Approach}} (\bibinfo{publisher}{Cambridge University Press},
  \bibinfo{year}{2004}).

\bibitem[{\citenamefont{NLANR}()}]{NLANR}
\bibinfo{author}{\bibnamefont{NLANR}}, \bibinfo{howpublished}{National
  Laboratory for Applied Network Research, http://moat.nlanr.net/}.

\bibitem[{\citenamefont{CAIDA}()}]{CAIDA}
\bibinfo{author}{\bibnamefont{CAIDA}}, \bibinfo{howpublished}{Cooperative
  Association For Internet Data Analysis, http://www.caida.org/}.

\bibitem[{ore()}]{oregon}
\bibinfo{howpublished}{Route Views Project. University of Oregon, Eugene.
  http://www.routeviews.org/}.

\bibitem[{mic()}]{michigan}
\bibinfo{howpublished}{Topology Project, University of Michigan, Ann Arbor.
  http://topology.eecs.umich.edu/}.

\bibitem[{ski()}]{skitterdata}
\bibinfo{howpublished}{The Data Kit \#0204 was collected as part of CAIDA's
  Skitter initiative. Support for Skitter is provided by
  DARPA, NSF, and CAIDA membership.}

\bibitem[{\citenamefont{Barab\'asi and Albert}(1999)}]{barabasi99a}
\bibinfo{author}{\bibfnamefont{A.~L.} \bibnamefont{Barab\'asi}}
  \bibnamefont{and} \bibinfo{author}{\bibfnamefont{R.}~\bibnamefont{Albert}},
  \bibinfo{journal}{Science} \textbf{\bibinfo{volume}{286}},
  \bibinfo{pages}{509} (\bibinfo{year}{1999}).

\bibitem[{\citenamefont{Barab\'asi et~al.}(1999)\citenamefont{Barab\'asi,
  Albert, and Jeong}}]{barabasi99b}
\bibinfo{author}{\bibfnamefont{A.~L.} \bibnamefont{Barab\'asi}},
  \bibinfo{author}{\bibfnamefont{R.}~\bibnamefont{Albert}}, \bibnamefont{and}
  \bibinfo{author}{\bibfnamefont{H.}~\bibnamefont{Jeong}},
  \bibinfo{journal}{Physica A} \textbf{\bibinfo{volume}{272}},
  \bibinfo{pages}{173} (\bibinfo{year}{1999}).

\bibitem[{\citenamefont{Zhou and Mondrag\'on}(2004{\natexlab{a}})}]{zhou04a}
\bibinfo{author}{\bibfnamefont{S.}~\bibnamefont{Zhou}} \bibnamefont{and}
  \bibinfo{author}{\bibfnamefont{R.~J.} \bibnamefont{Mondrag\'on}},
  \bibinfo{journal}{IEEE Comm. Lett.} \textbf{\bibinfo{volume}{8}},
  \bibinfo{pages}{180} (\bibinfo{year}{2004}{\natexlab{a}}).

  \bibitem[{\citenamefont{Newman}(2002)}]{newman02}
\bibinfo{author}{\bibfnamefont{M.~E.~J.} \bibnamefont{Newman}},
  \bibinfo{journal}{Phys. Rev. Lett.} \textbf{\bibinfo{volume}{89}},
  \bibinfo{pages}{208701} (\bibinfo{year}{2002}).

\bibitem[{\citenamefont{Newman}(2003)}]{newman03}
\bibinfo{author}{\bibfnamefont{M.~E.~J.} \bibnamefont{Newman}},
  \bibinfo{journal}{Phys. Rev. E} \textbf{\bibinfo{volume}{67}},
  \bibinfo{pages}{026126} (\bibinfo{year}{2003}).

\bibitem[{\citenamefont{Maslov et~al.}(2004)\citenamefont{Maslov, Sneppen, and
  Zaliznyak}}]{maslov04}
\bibinfo{author}{\bibfnamefont{S.}~\bibnamefont{Maslov}},
  \bibinfo{author}{\bibfnamefont{K.}~\bibnamefont{Sneppen}}, \bibnamefont{and}
  \bibinfo{author}{\bibfnamefont{A.}~\bibnamefont{Zaliznyak}},
  \bibinfo{journal}{Physica A} \textbf{\bibinfo{volume}{333}},
  \bibinfo{pages}{529} (\bibinfo{year}{2004}).



\bibitem[{\citenamefont{Zhou and Mondrag\'on}(2004{\natexlab{b}})}]{zhou04b}
\bibinfo{author}{\bibfnamefont{S.}~\bibnamefont{Zhou}} \bibnamefont{and}
  \bibinfo{author}{\bibfnamefont{R.~J.} \bibnamefont{Mondrag\'on}},
  \bibinfo{journal}{IEE Elec. Lett.} \textbf{\bibinfo{volume}{40}},
  \bibinfo{pages}{151} (\bibinfo{year}{2004}{\natexlab{b}}).

\bibitem[{\citenamefont{Dorogovtsev and Mendes}(2000)}]{dorogovtsev00}
\bibinfo{author}{\bibfnamefont{S.~N.} \bibnamefont{Dorogovtsev}}
  \bibnamefont{and} \bibinfo{author}{\bibfnamefont{J.~F.~F.}
  \bibnamefont{Mendes}}, \bibinfo{journal}{EuroPhys. Lett.}
  \textbf{\bibinfo{volume}{52}}, \bibinfo{pages}{33} (\bibinfo{year}{2000}).

\bibitem[{\citenamefont{Bu and Towsley}(2002)}]{bu02}
\bibinfo{author}{\bibfnamefont{T.}~\bibnamefont{Bu}} \bibnamefont{and}
  \bibinfo{author}{\bibfnamefont{D.}~\bibnamefont{Towsley}}, in
  \emph{\bibinfo{booktitle}{Proc. of IEEE INFOCOM 2002}}
  (\bibinfo{year}{2002}), p. \bibinfo{pages}{638}.

\bibitem[{\citenamefont{Bianconi et~al.}(2003)\citenamefont{Bianconi,
  Caldarelli, and Capocci}}]{bianconi03}
\bibinfo{author}{\bibfnamefont{G.}~\bibnamefont{Bianconi}},
  \bibinfo{author}{\bibfnamefont{G.}~\bibnamefont{Caldarelli}},
  \bibnamefont{and} \bibinfo{author}{\bibfnamefont{A.}~\bibnamefont{Capocci}},
  \emph{\bibinfo{title}{Number of h-cycles in the Internet at the autonomous
  system level}}, \bibinfo{howpublished}{ArXiv:cond-mat/0310339}
  (\bibinfo{year}{2003}).

\bibitem[{\citenamefont{Caldarelli et~al.}(2004)\citenamefont{Caldarelli, Rios,
  and Pietronero}}]{caldarelli03}
\bibinfo{author}{\bibfnamefont{G.}~\bibnamefont{Caldarelli}},
  \bibinfo{author}{\bibfnamefont{P.~D.~L.} \bibnamefont{Rios}},
  \bibnamefont{and}
  \bibinfo{author}{\bibfnamefont{L.}~\bibnamefont{Pietronero}},
  \emph{\bibinfo{title}{Generalized network growth: from microscopic strategies
  to the real Internet properties}},
  \bibinfo{howpublished}{arXiv:cond-mat/0307610 v1} (\bibinfo{year}{2004}).

\bibitem[{\citenamefont{Zhou and Mondrag\'on}(2003)}]{zhou03c}
\bibinfo{author}{\bibfnamefont{S.}~\bibnamefont{Zhou}} \bibnamefont{and}
  \bibinfo{author}{\bibfnamefont{R.~J.} \bibnamefont{Mondrag\'on}}, in
  \emph{\bibinfo{booktitle}{Proc. of 18 Int. Teletraffic Congress (ITC18)}},
  edited by \bibinfo{editor}{\bibfnamefont{J.}~\bibnamefont{Charzinski}}
  (\bibinfo{publisher}{Elsevier}, \bibinfo{address}{Berlin, German},
  \bibinfo{year}{2003}), vol.~\bibinfo{volume}{5a} of
  \emph{\bibinfo{series}{Teletraffic Science and Engineering}}, pp.
  \bibinfo{pages}{121--130}.

\bibitem[{\citenamefont{Krapivsky et~al.}(2000)\citenamefont{Krapivsky, Redner,
  and F.Leyvraz}}]{krapivsky00}
\bibinfo{author}{\bibfnamefont{P.~L.} \bibnamefont{Krapivsky}},
  \bibinfo{author}{\bibfnamefont{S.}~\bibnamefont{Redner}}, \bibnamefont{and}
  \bibinfo{author}{\bibnamefont{F. Leyvraz}}, \bibinfo{journal}{Phys. Rev.
  Lett.} \textbf{\bibinfo{volume}{85}}, \bibinfo{pages}{4629}
  (\bibinfo{year}{2000}).

\bibitem[{\citenamefont{Watts}(1999)}]{watts99}
\bibinfo{author}{\bibfnamefont{J.}~\bibnamefont{Watts}},
  \emph{\bibinfo{title}{Small Worlds: The Dynamics of Networks between Order
  and Randomness}} (\bibinfo{publisher}{Princeton Univeristy Press},
  \bibinfo{address}{New Jersey, USA}, \bibinfo{year}{1999}).

\bibitem[{\citenamefont{Bianconi and Capocci}(2003)}]{bianconi03a}
\bibinfo{author}{\bibfnamefont{G.}~\bibnamefont{Bianconi}} \bibnamefont{and}
  \bibinfo{author}{\bibfnamefont{A.}~\bibnamefont{Capocci}},
  \bibinfo{journal}{Phys. Rev. Lett.} \textbf{\bibinfo{volume}{90}},
  \bibinfo{pages}{078701} (\bibinfo{year}{2003}).

\bibitem[{\citenamefont{Goh et~al.}(2001)\citenamefont{Goh, Kahng, and
  Kim}}]{goh01}
\bibinfo{author}{\bibfnamefont{K.~I.} \bibnamefont{Goh}},
  \bibinfo{author}{\bibfnamefont{B.}~\bibnamefont{Kahng}}, \bibnamefont{and}
  \bibinfo{author}{\bibfnamefont{D.}~\bibnamefont{Kim}},
  \bibinfo{journal}{Phys. Rev. Lett.} \textbf{\bibinfo{volume}{87}},
  \bibinfo{pages}{278701} (\bibinfo{year}{2001}).

\bibitem[{\citenamefont{Holme and Kim}(2002)}]{Holme02a}
\bibinfo{author}{\bibfnamefont{P.}~\bibnamefont{Holme}} \bibnamefont{and}
  \bibinfo{author}{\bibfnamefont{B.~J.} \bibnamefont{Kim}},
  \bibinfo{journal}{Phys. Rev. E} \textbf{\bibinfo{volume}{65}},
  \bibinfo{pages}{066109} (\bibinfo{year}{2002}).

\bibitem[{\citenamefont{Hyun et~al.}()\citenamefont{Hyun, Broido, and
  claffy}}]{hyun03}
\bibinfo{author}{\bibfnamefont{Y.}~\bibnamefont{Hyun}},
  \bibinfo{author}{\bibfnamefont{A.}~\bibnamefont{Broido}}, \bibnamefont{and}
  \bibinfo{author}{\bibfnamefont{k.}~\bibnamefont{claffy}},
  \bibinfo{note}{http://www.caida.org/outreach/papers/2003/ASP/}.

\end{thebibliography}
\end{document}